# Developing and deploying deep learning models in brain MRI: a review

Kunal Aggarwal[1,2], Marina Manso Jimeno[3,4], Keerthi Sravan Ravi[3,4], Gilberto Gonzalez[5], Sairam Geethanath[1]

[1] Accessible MR Laboratory, Biomedical Engineering, and Imaging Institute, Dept. of Diagnostic, Molecular and Interventional Radiology, Mount Sinai Hospital, New York City, New York

[2] Department of Electrical and Computer Engineering, Technical University Munich, Munich, Germany

[3] Department of Biomedical Engineering, Columbia University in the City of New York, New York City, New York, USA

[4] Columbia University Magnetic Resonance Research Center, Columbia University in the City of New York, New York City, New York, USA

[5] Division of Neuroradiology, Department of Radiology, Massachusetts General Hospital, Boston, Massachusetts

Word Count: 5952

## **Abstract**

Magnetic Resonance Imaging (MRI) of the brain has benefited from deep learning (DL) to alleviate the burden on radiologists and MR technologists, and improve throughput. The easy accessibility of DL tools have resulted in the rapid increase of DL models and subsequent peer-reviewed publications. However, the rate of deployment in clinical settings is low. Therefore, this review attempts to bring together the ideas from data collection to deployment into the clinic building on the guidelines and principles that accreditation agencies have espoused. We introduce the need for and the role of DL to deliver accessible MRI. This is followed by a brief review of DL examples in the context of neuropathologies. Based on these studies and others, we collate the prerequisites to develop and deploy DL models for brain MRI. We then delve into the guiding principles to practice good machine learning practices in the context of neuroimaging with a focus on explainability. A checklist based on the FDA's good machine learning practices is provided as a summary of these guidelines. Finally, we review the current challenges and future opportunities in DL for brain MRI.

**Abbreviations:**

CAD: Computer Aided Detection

CAM: Class Activation Mapping

CNN: Convolutional Neural Network

DAGMNet: Dual attention gate network

DICOM: Digital Imaging and Communications in Medicine

DSC: DICE Similarity Coefficient

FL: Federated Learning

GDPR: General Data Protection Regulation

GMLPs: Good Machine Learning Practices

Grad-CAM: Gradient-weighted Class Activation Mapping

HR: High-Resolution

IRB: Institutional Review Board

LR: Low-Resolution

MCI: Mild Cognitive Impairment

OECD: Organisation for Economic Co-operation and Development

OOD: Out-of-Distribution

PPML: Privacy Protecting Machine Learning

PSNR: Peak Signal-to-Noise Ratio

RF: Radio Frequency

SSIM: Structural Similarity Index Measure

XAI: Explainable Artificial Intelligence

## Introduction

As of 2018[1-2], the density of MR scanners was least in geographies with the highest populations such as in sub-Saharan Africa and the Indian subcontinent. The severe lack of neurosurgeons and skilled human resources required to operate, use, and interpret data from MR scanners is a major barrier to accessing this life-saving technology[2]. In contrast, contemporary radiology departments in the Organisation for Economic Co-operation and Development (OECD) countries require a close interplay of a team with diverse expertise: radiologists specializing in different anatomical sites, MR technologists, medical physicists, and radiologic nurses supported by the vendor's service and application engineers. This MR inaccessibility and the resulting disparity necessitates the development and deployment of automated methods to augment existing local expertise. Recently, there has been the development of autonomous MRI methods to automate protocolling[3-4], identify artifacts[5-6], and reconstruct images from accelerated scans[7]. These advances are expected to assist local MR technicians, accelerate acquisitions, improve throughput by assisting or replacing manual data processing steps, and reduce waiting times for radiology reporting. In addition to positively impacting MR accessibility, these automation methods facilitate MRI-based big data studies such as the Human Connectome Project[8], UK Biobank[9], and Rhineland study[10], among others. This is due to the high volume and velocity associated with such studies.

These automation methods leverage the recent resurgence of artificial intelligence techniques in general and supervised learning methods in particular. A deep cascade of neural networks - termed deep learning (DL) models - is trained with a set of input features on one side and the resulting outcomes (labels) on the other side, using existing apriori annotated data[11]. These trained DL models are then validated against an unseen but similar data set to fine-tune "hyperparameters" of the DL model. Subsequently, the tuned DL model is used to perform inference on a test dataset to evaluate its performance by comparing it with a human-evaluated outcome. Finally, once the classification-tasked model performs satisfactorily with respect to false-positive (FP), false-negative (FN), true-positive (TP), and true-negative (TN) prediction metrics (captured in a "confusion matrix"), then it is considered for a deployment study and downstream accreditation processes[11].

In this review, we discuss studies that have developed and deployed deep learning models tasked with multi-task classification. We then present an analysis of the related literature to provide critical steps involved in data collection and curation required to set up deep learning studies. We then highlight the importance of good machine learning practices and the explainability of the DL results by illustrating examples and tools. A suggestive set of steps to enable mounting a successful development and deployment of MRI DL models will be then discussed based on recent literature. Finally, a brief overview of the challenges and opportunities related to MRI-based DL studies is presented.

## MRI DL studies of the brain

Easy and direct availability of vast amounts of MRI data from publicly available repositories such as HCP[8], UK Biobank[9], among others, as well as accessible tools to build[12] and optimize DL models[13] have significantly accelerated the application of DL methods to address challenges in MRI. These studies have resulted in a substantial body of peer-reviewed literature (Figure 1), with most of them sharing an open-source implementation of their models with sample data. This review focuses on MRI DL studies that meet the following criteria: (i) published in the last five years; (ii) Methods mostly focusing on classification and not regression tasks; (iii) studies incorporating explainable AI components; (iv) works demonstrating deployment of the DL models and preferably across multiple vendors and sites. These criteria were used to focus our review on specific developments. Notably, the generic tools developed by mathematicians and computer scientists in the DL community for explainable AI such as GradCAM[14] and other methods[15-16] have focused more on classification tasks compared to regression tasks (Figure 1). The quantification of the outcomes of a classifier network is relatively straightforward compared to a regression model. The outcomes are directly compared to the "true annotated labels" and hence result in a binary or a multi-class decision. These can easily be binned into TP, FP, TN, and FN and evaluated for sensitivity and specificity. In contrast, regression models accomplishing tasks such as DL denoising and synthesizing images are quantified using peak signal-to-noise ratio (PSNR) and structural similarity index measure (SSIM). These metrics have a continuous value and are hard to set thresholds for acceptance. In addition, regression models are typically explained using activation maps that are subsequently interpreted by the authors rather than community-wide tools independent of the developed methods. The interested reader is referred to [7] for a detailed review of machine learning-based image reconstruction methods that leverage regression models.

**Example DL MRI solutions for neuroimaging**

**Brain tumors:** Nalepa et al. have demonstrated a fully automated pipeline for DCE-MRI analysis of brain tumors called Sens.AI DCE[17]. In particular, they have substituted manual segmentation of brain tumors in $T_2$ FLAIR images with deep-learning-based methods to demonstrate improved reproducibility. They complement this with a real-time image processing algorithm to determine the vascular input region. Finally, they include a new cubic model of the vascular function for PK modeling. They have validated their package with the BraTs dataset for DICE coefficient and area under the curve as well as by twelve readers from two institutions. These results show good to excellent agreement between the gold standard BraTS dataset and Sens.AI DCE with a total execution time of approximately 3 minutes. Another study focused on the automated identification and classification of brain tumor MRI data classified into glioma, meningioma, and pituitary tumors with an accuracy of 93.7% and a DICE similarity coefficient (DSC) of 95.8%[18]. A subsequent task of classifying gliomas into high or low grade had an accuracy of 96.5% and a DSC of 94.3%. These models were based on the GoogleNet variant architectures to efficiently combine local features. The identification and classification tasks were accomplished in less than 3 minutes. The method compared well with other state-of-the-art methods with respect to accuracy. Although the study did not explicitly focus on explainable AI methods such as Grad-CAM for interpretation, the authors performed an ablation study to demonstrate the effect of the locally chosen features. Brain extraction is a key step in multiple neuroimaging pre-processing pipelines and in complying with privacy laws. This task becomes more challenging in the presence of pathology such as diffuse gliomas as most analytical and deep learning methods focus on healthy brain extraction. Thakur et al have addressed this gap by developing and testing their models for brain extraction in the presence of diffuse glioma, in a multi-institutional manner[19]. The authors considered multiparametric MRI data from private and public repositories acquired with different acquisition protocols to train a "modality-agnostic" tool that does not require retraining. The work demonstrated a similar or better accuracy compared to other brain extraction models that worked only on healthy brain extractions. The interested reader is pointed to these reviews for further reading on deep learning methods for brain tumor imaging[20], classification[21], and segmentation[22].

**Stroke:** The need for automated segmentation and classification of images, especially in emergency room settings in this time-critical pathology is well understood. In this direction, Liu et al developed a deep learning model that detected and segmented abnormalities in acute ischemic stroke[23]. The work included several steps of pre-processing the data, such as skull stripping and DWI intensity normalization, among others. The study compared T-score and the modified c-fuzzy methods for lesion segmentation. In addition, the authors implemented the 3D Dual attention gate network (DAGMNet) as a supervised learning method to delineate the lesions. The developed model performs better than the unsupervised generic tools and is faster, publicly available, and easy to deploy. They tested their algorithm on a hold-out data set of 280 MRIs and quantified the improved performance using DICE scores, precision, sensitivity, subject detection rate, DICE scores for the lesion volumes and lesion DWI contrast. Another study on stroke detection performed by Zhang et al demonstrated an accuracy of 89.77% over 300 ischemic stroke patients[24]. The authors evaluated three network architectures with labels drawn by experienced radiologists from two hospitals. Their models predicted a bounding box covering the lesions. Statistical analysis performed on the location, size, and shape correlated well with the radiologists' labeling. This implementation aids in rapid localization and preliminary characterization of the lesion. The authors have committed to making the data available once a thousand patient data are collected. An important task in imaging stroke is to grade the severity of the ischemia. A multi-class classification solution for detecting the severity has been developed by Acharya et al.[25] The authors extracted higher-order features from MR images, such as bispectrum entropy and its phase, followed by support vector machines to classify the severity of the stroke into LACS, PACS and TACS. The algorithm demonstrated high levels of accuracy without the need for any manual intervention to augment the neuroradiologist.

**Alzheimer's disease:** Supervised learning models have demonstrated the utility of automation in the imaging of Dementia. A specific challenge in this area is the classification and staging of the progression in mild cognitive impairment (MCI) patients. Kwak et al. have developed a deep learning model based on brain atrophy patterns and associated these changes with differences in amyloid burden, cognition, and metabolism[26]. This model is used to classify AD patients from cognitively normal subjects. A secondary classification helps identify the trajectory of cognitive decline in individuals with MCI. These results were validated with cognitive tests, fluid biomarkers, and PET uptake data with good agreement. The approach is expected to benefit from integrating cognitive and neurobiological features to capture the heterogeneity of MCI. Another approach involved the development and testing of whole-brain 3D convolutional neural networks to detect

AD[27]. This model did not include any patient-specific information to allow the generalization of the algorithm. The implementation included four steps of brain extraction, normalization, 3D CNN followed by domain adaptation. A key feature of this study is the authors' focus on accomplishing accountability. The method outperformed other state-of-the-art methods in the CAD Dementia challenge test, along with explainable AI visualizations to aid the interpretation of classification results. Ahmed et al. also used a 3D approach but collected an ensemble of 2D patches in three orientations to train an ROI-based neural network to stage AD using MR images as per NIA labels[28]. This approach was demonstrated on the GARD and ADNI datasets. This method was compared to other state-of-the-art methods, and the performance was similar or better. The landmarks delineated by the algorithm to indicate AD correlated well with known neuroanatomical areas. However, the model could not accurately predict asymptomatic AD based on the high rates of false positives.

**Prerequisites for DL-based neuro-MRI**

Data from routine MRI studies result in high volume, velocity, and variety: characteristics of big data[29–35]. For training DL models, high volume and velocity are favorable factors: more data is better than lesser; high velocity of data requires automated processing methods. However, the variety in MRI data due to a large number of available acquisition parameters, reconstruction methods, receive coil configurations, post-processing steps requires attention to fine details before collating data for annotation and subsequent training. In line with the classification suggested by Wald[36], the sources of these variabilities need to be binned as emanating from the MR system characteristics, subject-induced variations, and pathology-specific factors. Typically, the goal of the DL models discussed in this work is to glean the subtle changes in pathophysiological states based on the MR images. Therefore, standardization of parameters getting affected by the system and subject priors is critical to ensure that the DL models focus their attention on the pathology being investigated. The removal or reduction of the confounding variables is therefore essential in building these DL models. This exercise is facilitated by the use of explainable AI tools. Therefore, two critical requirements to understand an MRI-based DL study are standardization of procedures and methods and explainable AI tools. A detailed discussion on these prerequisites is performed below.

DL models require large amounts of data to achieve high accuracy levels[37–39]. Unlike models trained on natural images, large datasets are challenging to achieve for medical imaging applications[37–39]. Training a DL model of medical images entails data collection, curation, and annotation. Additionally, data augmentation might be necessary in cases when the data are insufficient or to strengthen the generalization ability of the model[39–41]. The steps of data collection and annotation are the most expensive and time-consuming, and patient privacy policies often restrict the use and sharing of images[37,39,42,43]. These factors significantly impact the models' performance in clinical settings, limiting their ultimate deployment[40,42]. The purpose of this section is to review strategies and recommendations (Figure 2) at the data level for maximizing the likelihood of successful deployment after training based on previously published work.

Data collection, including patient selection, imaging protocol, sequences, and scan parameters, is determined based on the intended use of the model and its targeted application. Cohorts can be prospective or retrospective, depending on whether the data are acquired for the study or retrieved from a public or private repository[44]. An initial step in the process is the approval by an Institutional Review Board (IRB) or a similar board. Additionally, participants provide informed consent about the use of their personal data, which is typically de-identified during data curation. Privacy Protecting Machine Learning (PPML) is a niche area of research that aims at maximizing the confidentiality of patient data while optimizing its use on data-driven models[45]. In this field, Federated Learning (FL) alleviates the shortage of data problems by allowing training models on large-scale, multi-center data without data sharing. FL feasibility in MRI has been explored by Sarma et al. and Sheller et al. for brain tumor and prostate segmentation tasks[46,47]. When dealing with multi-institutional data, protocol harmonization can avoid model bias that may arise from differences in image contrast, intensity, or noise distribution. In MRI, these differences may stem from multiple sources, including variations in system manufacturer, field strength, Radio Frequency (RF) coils, patient positioning, acquisition sequence and scan time, and even pre-processing and reconstruction pipelines.

Publicly available databases are the results of extensive research projects and contain large amounts of data that can be leveraged for training. These datasets are typically acquired using the same protocol and scanner or using highly harmonized protocols. Patient inclusion and exclusion criteria in these research cohorts are strict, and the datasets are well-curated and usually undergo multi-step post-processing pipelines and standardization operations. While the reproducibility of models trained on publicly available datasets is easier to assess, the data lack the heterogeneity characteristic of clinical data observed during deployment. Martensson et al.[43] systematically studied the performance variability of a DL model trained on different combinations of training sets, including publicly available datasets and more heterogeneous Out-of-Distribution (OOD) datasets. They observed that performance drops when models trained on homogeneous data are applied to clinical data. However, inference on clinical data showed a better agreement level with a radiologist reading if clinical cohorts were present in the training data.

A benefit of local data acquisition for training is having access to raw data. Most public or private imaging repositories contain data in the Digital Imaging and Communications in Medicine (DICOM) format. The raw data undergo several pre-processing steps, including coil-combination, filtering, artifact correction, and phase removal before storage, stripping the images of features that DL models in the process could recognize. Raw data might be favored for certain DL tasks, data augmentation techniques, or for data synthesis via forward modeling. This is exemplified by the increasing usage of the fastMRI dataset[48], the only publicly-available dataset of raw MR knee and brain data. It has become a benchmark for the validation and reproducibility assessment of DL-based image reconstruction algorithms. Additionally, models for the detection or correction of k-space-occurring artifacts such as motion[49–51] and Gibbs ringing[5,52] typically leverage raw data for the simulation of artifact-corrupted images.

Data collection is followed by data curation. This step is performed to standardize and improve dataset quality for subsequent deep neural network training[42]. The data used for the model development entails a trade-off between distribution heterogeneity and representation bias. It should represent varying patient populations and anatomy disparities while avoiding biasing network representation. Successfully deployed models are typically developed with data acquired using the same imaging protocol and the same system as the site targeted for deployment[53,54]. Failure mode analysis of the prospective evaluation of an automatic kidney segmentation model after deployment revealed segmentation errors arising from common clinical scenarios such as a fluid-filled stomach and a distended bladder[54]. These cases are typically excluded during cohort building or data curation and reduce the model's tolerance to data variations. Poor generalizability to unseen domains is one of the major challenges to successfully deploying DL models in the clinic[55].

For fully and semi-supervised learning tasks, data labeling or annotation is typically the most time-consuming step of an AI project. It may require localizing, delineating, or segmenting lesions or organs of interest or labeling or annotating characteristics of the data. This step is performed manually by an experienced reader via visual inspection. When human observations or clinicians' expertise is required to annotate the data, multiple readers and ideally with variable levels of experience, are preferred to estimate inter-reader variability and compare it to the model's performance.

Finally, data augmentation is the process of generating additional versions of the initial data to enlarge the training set and improve the model's robustness, thereby avoiding overfitting[56]. Typical techniques include translation, flipping, rotation, and cropping. Depending on the application, other approaches may be useful, such as random k-space oversampling for model-based reconstruction techniques. Data augmentation can also be leveraged to reduce the gap between the training data and prospective clinical data, for example, by simulating noise and motion artifacts in the images. A recent study[57] for accelerated MR reconstruction demonstrated that with the introduction of these commonly-occurring artifacts using MR physics-driven data augmentation techniques, model performance on both in-distribution and OOD data increases compared to state-of-the-art image-based data augmentation methods.

**Good Machine Learning Practices for MRI**

AI's novelty and current regulatory paradigms are not well adapted to strike a balance between patient safety and promoting the expansion of this new industry[58]. For assessing commercially accessible algorithms to guarantee their dependability and safety, defining best practices is an area of active research[59,60], significant regulatory problems need to be resolved to move clinical AI toward becoming safe and robust[61]. Wu et. al. reviewed the FDA database for parameters that were used to evaluate the AI algorithms of products. The parameters they found were - (i) number of patients and sites used in the evaluation, (ii) prospective and retrospective collection of data, and (iii) whether the performance was stratified by disease subtypes or not. Based on the FDA summary, the study revealed that 126 out of 130 AI devices conducted solely retrospective investigations at their submission. The influence of the AI decision tool on clinical practice must be fully characterized, though, and this is crucial since human-computer interaction might differ significantly from a model's intended purpose. For instance, most computer-aided detection (CAD) diagnostic tools are meant to serve as decision-support aids rather than primary diagnostic instruments[61]. Instead of independently diagnosing, staging, or triaging pathology, CAD is meant to identify, mark, highlight, or otherwise draw attention to imaging characteristics[62]. The FDA suggests regulating AI software based on function rather than technical components or intended use, which is different from the case for most pharmaceutical items, gadgets, and foods[58].

Therefore, FDA advocates ten guiding principles for medical device development known as 10 Good Machine Learning Practices (GMLPs) that take into consideration the prerequisites discussed above. According to the first and second principle, all expertise related to the product development should work together from the development phase until integration into the clinical workflow. This includes neuroradiologists, neuroimaging scientists, MR technicians, and data scientists implementing good software engineering and security practices. The third principle states the importance of metadata in developing DL models and connects to the concept of data security from the second principle. The fourth and fifth principles mention the importance of datasets. The training and testing datasets should be independent and the reference datasets should have the same characteristics as of the patients in the former datasets. According to the sixth principle, the intended use of a model must be clearly defined along with its risks and performance limitations on different datasets. This relates to principle number three in a sense that metadata defines the scope of the model being used on the specific patients. Principle seven states the involvement of humans and the fact that human intervention cannot be avoided at any stage of development or deployment. This principle focuses more on AI in the loop rather than human in the loop. Eighth and ninth principle centers on the user and states that the model should be easy to understand for the end user and must list all the possible precautions in order to avoid harm to the patient. Principle ten mentions that updates in the model are a mandatory part of the DL deployment and must be considered frequently[59]. A checklist based on these GMLPs is provided as a summary specifically designed for experts working in brain MRI (Table 1).

**Explainable AI**

Although deep learning techniques produce outcomes, they do not explain how those results were obtained. One cannot just analyze the deep neural network to understand how that choice was made. As a result, deep learning models are sometimes referred to as "Black Boxes"[63]. Medical professionals believe these "black boxes" may be prejudiced in some way, which might have negative effects when used in practical applications[63]. Additionally, laws like the General Data Protection Regulation (GDPR, Article 15) of the European Union specify that patients have the right to request an explanation for how a given diagnosis was reached if the standard deep learning models cannot[64]. Therefore, Explainable Artificial Intelligence (XAI) techniques recently developed with the primary objective of visualizing and interpreting the results of machine learning (ML) and deep learning (DL) networks represent a potential remedy to close this gap between high performance and deep-level understanding[65] (Figure 3). They have been utilized in a variety of applications, including the categorization of ECGs[66] and the visualization of feature maps at various Convolutional Neural Network (CNN) layers[67].

Velden et al. categorized XAI approaches into three groups based on three criteria: (i) model-based vs post-hoc; (ii) model-specific against model-agnostic; and (iii) global versus local. These categories are visual, textual, and example based. The most prevalent type of XAI in medical imaging, out of these three categories, is the visual explanation. These approaches, sometimes referred to as saliency mapping, employ a backpropagation methodology to highlight the key elements of a picture for a certain model's decision by emphasizing the pixels that had the greatest influence on the results of the investigation[64]. Class activation mapping (CAM), a technique used in the backpropagation methodology, was introduced by Zhou et al. in 2016. They used global average pooling on the last convolutional feature maps to substitute the fully connected layers at the conclusion of a CNN. It is a weighted linear sum of the visual patterns that were observed and recorded by the filters at various spatial positions[68].

Gradient-weighted class activation mapping is a generic strategy that includes CAM as one of its specialized methods (Grad-CAM). Grad-CAM can operate with any CNN, but CAM needs global average pooling in particular[64]. Grad-CAM delivers the ROI on an input image that has the greatest influence on class prediction. Grad-CAM allows us to track the spatial attention changes that occur between network layers, or more precisely, what each network layer focuses on in each input image. To do this, the output gradient with respect to each neuron in the network is calculated to ascertain its relative significance[69].

Grad-CAM has been widely used to describe deep learning models. Jimeno et. al. used it to identify and classify wrap-around and Gibbs ringing artifacts[5]. It was used by Windisch et al. in 2020 to identify brain MRI regions that caused the classifier to determine the existence of a malignancy[70]. A model's prediction of the fetus's brain age may also be explained using Grad-CAM, according to a 2020 publication by Liao et al., which will help avoid congenital malformations[71]. It was also employed by Natekar et al. in 2020 to describe the brain tumor segmentation network[69].

Occlusion Sensitivity technique is another XAI tool[64]. The input MR image is disturbed by a small perturbation, and the categorization choice is changed and examined. In order to quantify the variation in the output prediction, it covers a piece of the input picture with a black patch. After moving the patch across the whole picture, it is simple to determine which parts of the brain are responsible for the categorization choice in question by looking at this variation[65]. This approach was utilized by Bordin et al. to identify relationships between White matter hyperintensities and the anatomical areas that are most important for the categorization of Alzheimer's disease. In conclusion, these XAI approaches present a potentially important addition that may eventually boost radiologist's confidence in the usage of AI models.

## Challenges and opportunities

DL research has recently witnessed accelerating adoption in the field of MRI (Figure 1) impacting image acquisition, reconstruction, processing, and radiological reporting tasks.

**Image acquisition:** Currently, DL for MRI acquisition can be classified into two broad categories: (i) automatically generating MR pulse sequences for a target contrast or signal-to-noise ratio. In this approach, once a vendor hardware of interest has been identified, imposing appropriate constraints on the cost function (for example, slew rate) will facilitate easy implementation of the optimized pulse sequence on the chosen hardware[4]. Second is the acceleration of existing vendor-defined protocols, potentially relying on post-acquisition methods to recover SNR[72–74]. This approach is inherently limited to a particular protocol and vendor. The emergence of physics-informed DL methods will allow researchers to develop models that are privy to the underlying physical phenomena, potentially resulting in improved interpretability since the outputs can be evaluated using existing task-specific knowledge[75–79]. Performing automated and intelligent slice planning for localizers is also an active area of research[80,81].

**Image reconstruction and processing:** Based on the work by Chaudhari et al.[40], applications of DL to image reconstruction and processing are classified into model-free image synthesis, model-based image reconstruction, and classification and segmentation. Model-free image synthesis pertains to the mapping of input images to output images. Examples are image super-resolution, denoising, artifact reduction or removal, and synthesis of missing contrasts. Image super-resolution enables the acquisition of multiple low-resolution images, which can be upscaled using DL models[82–87]. Compiling a training dataset for this task is not straightforward since it is not trivial to acquire paired low-resolution (LR) and high-resolution (HR) data. Apart from logistical challenges, image registration is a primary concern. It is therefore convenient to acquire HR images and subsequently perform retrospective downsampling to generate LR images. However, this does not faithfully replicate MRI encoding, and hence does not accurately represent real-world LR data. In some other cases, HR data is not readily available. One workaround is to leverage a self-supervised learning framework to synthesise low-resolution images from high-resolution data, thereby mitigating the requirement of image registration[88]. Image denoising models improve SNR post-acquisition[72,74]. Two common approaches to achieve image denoising are to either directly synthesise the denoised image, or to synthesise the residual from which the final denoised image can be obtained. In the first approach, the models are trained on pairs of noisy/clean images to optimize for image quality whilst avoiding blurring artifacts and retaining the anatomical structures present in the original image[89–92]. An alternative method is to obtain the final denoised image from the difference of the original input image and the predicted residual[93,94]. Next, artifact reduction or removal models improve image quality by partially or completely correcting MR image artifacts that might otherwise interfere with diagnosis or reduce image quality[52,95–97]. Finally, contrast-synthesis models enable performing a limited MR exam whilst still obtaining the same diagnostic information as from a comprehensive MR exam, by generating the missing contrasts[98,99]. They can also enable performing contrast-enhanced MR examinations with reduced dosages of the exogenous contrast agents[100,101]. Model-based image reconstruction involves transforming undersampled data into fully-sampled reconstructed images[102–105]. One primary challenge associated with image synthesis and reconstruction is *hallucination*[40]. This relates to the addition of features that are not present in the input image. Since the model's representations are learned implicitly, hallucinations typically tend to reflect the characteristics of the training dataset. The challenge of distinguishing true image signals from hallucinated signals is exacerbated in the task of contrast-synthesis. Existing explainable AI approaches applicable to other tasks are not amenable to image synthesis tasks. Consequently, mitigation strategies to avoid hallucinations are an active area of research in the broader DL community. For image reconstruction, embedding data-consistency steps into the reconstruction process is a viable strategy to mitigate hallucinations.

**Radiological reporting:** The typical workflow of a radiologist involves identifying, localizing, and characterizing the pathology of interest. This is labour-intensive, and recent DL implementations have attempted to alleviate this burden on the radiologist[106]. Examples range from predicting diagnosis from input images, to generating a text-based radiological report from input images. The superior performance of DL methods on identification and classification tasks lends itself to the automated detection of findings from acquired images. Furthermore, several works have also demonstrated a potential for automated interpretation of findings[107]. Finally, assisting clinical decision support systems could improve quality of care[108,109]. However, the attribution of clinical decisions that were assisted by DL systems is an unresolved problem. Along with other ethical and legal challenges such as those related to data sharing and bias (refer to sections on prerequisites and GMLP), these bottlenecks need to be addressed prior to a potential deployment in a real-world scenario. Wang et al. discuss the entire workflow of medical imaging: from tomographic raw data/features to reconstructed images and then extracted diagnostic features/readings[7].

Figure 4 briefly captures the broader challenges and opportunities associated with employing DL in medical imaging. In general, DL methods present other challenges and opportunities apart from the application-specific ones discussed above. First, the current state-of-the-art DL qualifies as *narrow intelligence* since it lacks global context[108]. This results in severe performance degradation when tackling out of distribution data (OOD). Furthermore, it is not trivial to identify whether unseen data is OOD[110]. This problem is exacerbated in diagnostic healthcare imaging because the generated data is heterogeneous, noisy, and incomplete. This can be attributed to the differences in vendor hardware and software, and the plethora of component configurations[111]. Second, the lack of interpretability of DL models does not allow clinical users to develop trust in the models' predictions, resulting in stymied adoption and deployment in healthcare. Third, training DL models to achieve the level of robustness necessary to handle this variety requires an ImageNet-like breakthrough in the medical imaging community at large[112]. Recent works such as RadImageNet[112] are encouraging, and can potentially facilitate such advancements. However, with ever-growing scales of data collection, the closely coupled and critical task of data curation grows in complexity, at least for supervised learning frameworks. This relates to the fifth challenge, which involves ensuring bias-free data curation. The performance of any DL model is directly dependent on the quality of the data it was trained on. To avoid any biases in the output which could potentially compound in downstream analyses, the training dataset has to be free of all confounding factors. Training DL models on large-scale datasets requires prohibitively expensive hardware setups to provide the required compute, coupled with extremely long training durations. Consequently, this time-, cost-, and resource-intensive workflow raises the development barrier thereby mostly limiting research efforts to well-funded organizations and institutions. However, the recently increasing availability of commercial cloud solutions by Amazon, Microsoft, Google, etc., unlocks cost-effective compute that is globally accessible. In addition to the pre-existing heterogeneity of the data, acquisition methods are constantly evolving, introducing another dimension of variability to the data. This will require deployed DL models to be capable of online training to avoid incorrect or irrelevant predictions, or potential misdiagnoses in downstream analyses when encountering OOD data. Any development workflow lag, regardless of the duration, will result in incorrect treatment planning until updated models are deployed. On the other hand, disengaging the models until newer versions are available will result in workflow interruptions and throughput degradation. Lastly, the ethical and legal uncertainties involved critically need to be resolved prior to any potential deployments. Different countries enforce different medical data custody laws, necessitating region-specific modifications to the DL tool and the data pipeline to ensure compliance. Most importantly, the ownership of a DL-assisted clinical decision is an open question. Despite these challenges, the ability to automate tasks such as image interpretation and diagnosis will alleviate the immense burden on healthcare providers, allowing them to focus on other important tasks whilst improving the quality of their work lives. Providing more accurate and timely diagnosis, reduced costs, increased efficiency, and tailored treatments to individual patients based on their specific characteristics and needs all result in improved patient outcomes. These are strong motivators to strategize immediate or near-future adoption of existing DL methods. Directing research efforts to explore opportunities and simultaneously addressing existing issues will aid in the wider adoption and improved realization of DL's potential. Along with addressing weaknesses and leveraging strengths, incorporating the GMLP principles (Section 4) across the development lifecycle of DL-assisted medical applications will aid in maximizing safety, efficiency, and quality during clinical deployment.

## Conclusion

Our literature review indicates an increase in DL models for brain MRI tasks related to the acquisition, reconstruction, image analysis, and reporting in the last five years across neuropathologies such as tumors, stroke, and Alzheimer's disease. These studies were summarized as a suggestive DL pipeline for brain MRI studies. Importantly, the proportion of studies that adhere to GMLP principles and contain XAI components are significantly low. This DL neuro-MRI GMLP checklist in this review is motivated by this gap and emanates from the ten-point guidelines espoused by the accreditation agencies for these principles tailored to brain MRI. Finally, our assessment of the opportunities and challenges in DL studies on brain MRI indicates that the inclusion of the GMLPs significantly reduces the challenges associated with cost, and lack of interpretability, bias in the training data among others (Figure 4). Overcoming these challenges will unlock the potential to improve multiple aspects of neuroimaging using MRI through the successful deployment of accreditation agency-approved DL models.

## References:

[1] World Health Organization. Global atlas of medical devices. Published online 2022. https://www.who.int/publications/i/item/9789240062207

[2] Geethanath S, Vaughan JT Jr. Accessible magnetic resonance imaging: A review. *J Magn Reson Imaging*. 2019;49(7):e65-e77.

[3] Ravi KS, Geethanath S. Autonomous magnetic resonance imaging. *Magnetic Resonance Imaging*. 2020;73:177-185. doi:10.1016/j.mri.2020.08.010

[4] Loktyushin A, Herz K, Dang N, Glang F, Deshmane A, Weinmüller S, et al. MRzero - Automated discovery of MRI sequences using supervised learning. *Magn Reson Med*. 2021;86(2):709-724.

[5] Manso Jimeno M, Ravi KS, Jin Z, Oyekunle D, Ogbole G, Geethanath S. ArtifactID: Identifying artifacts in low-field MRI of the brain using deep learning. *Magn Reson Imaging*. 2022;89:42-48.

[6] Ahmad A, Parker D, Dheer S, Samani ZR, Verma R. 3D-QCNet - A pipeline for automated artifact detection in diffusion MRI images. *Comput Med Imaging Graph*. 2022;103:102151.

[7] Wang G, Ye JC, Mueller K, Fessler JA. Image Reconstruction is a New Frontier of Machine Learning. *IEEE Trans Med Imaging*. 2018;37(6):1289-1296.

[8] Van Essen DC, Smith SM, Barch DM, Behrens TEJ, Yacoub E, Ugurbil K, et al. The WU-Minn Human Connectome Project: an overview. *Neuroimage*. 2013;80:62-79.

[9] Littlejohns TJ, Holliday J, Gibson LM, Garratt S, Oesingmann N, Alfaro-Almagro F, et al. The UK Biobank imaging enhancement of 100,000 participants: rationale, data collection, management and future directions. *Nat Commun*. 2020;11(1):2624.

[10] Lohner V, Lu R, Enkirch SJ, Stöcker T, Hattingen E, Breteler MMB. Incidental findings on 3 T neuroimaging: cross-sectional observations from the population-based Rhineland Study. *Neuroradiology*. 2022;64(3):503-512.

[11] Goodfellow I, Bengio Y, Courville A. *Deep Learning*. MIT Press; 2016.

[12] Kheir AMS, Ammar KA, Amer A, Ali MGM, Ding Z, Elnashar A. Machine learning-based cloud computing improved wheat yield simulation in arid regions. *Comput Electron Agric*. 2022;203:107457.

[13] Howard J, Gugger S. Fastai: A Layered API for Deep Learning. *Information*. 2020;11(2):108. doi:10.3390/info11020108

[14] Selvaraju RR, Cogswell M, Das A, Vedantam R, Parikh D, Batra D. Grad-cam: Visual explanations from deep networks via gradient-based localization. In: *Proceedings of the IEEE International Conference on Computer Vision*. ; 2017:618-626.

[15] Sarker MK. *Towards Explainable Artificial Intelligence (XAI) Based on Contextualizing Data with Knowledge Graphs*. PhD. Kansas State University; 2020.

[16] Rahman MM. *Deep Interpretability Methods for Neuroimaging*. PhD. Georgia State University; 2022.

[17] Nalepa J, Ribalta Lorenzo P, Marcinkiewicz M, Bobek-Billewicz B, Wawrzyniak P, Walczak M, et al. Fully-automated deep learning-powered system for DCE-MRI analysis of brain tumors. *Artif Intell Med*. 2020;102:101769.

[18] Haq EU, Jianjun H, Li K, Haq HU, Zhang T. An MRI-based deep learning approach for efficient classification of brain tumors. *Journal of Ambient Intelligence and Humanized Computing*. Published online 2021. doi:10.1007/s12652-021-03535-9

[19] Thakur S, Doshi J, Pati S, Rathore S, Sako C, Bilello M, et al. Brain extraction on MRI scans in presence of diffuse glioma: Multi-institutional performance evaluation of deep learning methods and robust modality-agnostic training. *Neuroimage*. 2020;220:117081.

[20] Shaver MM, Kohanteb PA, Chiou C, Bardis MD, Chantaduly C, Bota D, et al. Optimizing Neuro-Oncology Imaging: A Review of Deep Learning Approaches for Glioma Imaging. *Cancers* . 2019;11(6). doi:10.3390/cancers11060829

[21] Muhammad K, Khan S, Ser JD, Albuquerque VHC de. Deep Learning for Multigrade Brain Tumor Classification in Smart Healthcare Systems: A Prospective Survey. *IEEE Trans Neural Netw Learn Syst*. 2021;32(2):507-522.

[22] Magadza T, Viriri S. Deep Learning for Brain Tumor Segmentation: A Survey of State-of-the-Art. *J Imaging Sci Technol*. 2021;7(2). doi:10.3390/jimaging7020019

[23] Liu CF, Hsu J, Xu X, Ramachandran S, Wang V, Miller MI, et al. Deep learning-based detection and segmentation of diffusion abnormalities in acute ischemic stroke. *Commun Med*. 2021;1:61.

[24] Zhang S, Xu S, Tan L, Wang H, Meng J. Stroke Lesion Detection and Analysis in MRI Images Based on Deep Learning. *Journal of Healthcare Engineering*. 2021;2021:1-9. doi:10.1155/2021/5524769

[25] Acharya UR, Rajendra Acharya U, Meiburger KM, Faust O, Koh JEW, Oh SL, et al. Automatic detection of ischemic stroke using higher order spectra features in brain MRI images. *Cognitive Systems Research*. 2019;58:134-142. doi:10.1016/j.cogsys.2019.05.005

[26] Kwak K, Giovanello KS, Bozoki A, Styner M, Dayan E, Alzheimer's Disease Neuroimaging Initiative. Subtyping of mild cognitive impairment using a deep learning model based on brain atrophy patterns. *Cell Rep Med*. 2021;2(12):100467.

[27] Folego G, Weiler M, Casseb RF, Pires R, Rocha A. Alzheimer's Disease Detection Through Whole-Brain 3D-CNN MRI. *Front Bioeng Biotechnol*. 2020;8:534592.

[28] Ahmed S, Kim BC, Lee KH, Jung HY, Alzheimer's Disease Neuroimaging Initiative. Ensemble of ROI-based convolutional neural network classifiers for staging the Alzheimer disease spectrum from magnetic resonance imaging. *PLoS One*. 2020;15(12):e0242712.

[29] Santos DP dos, dos Santos DP, Baeßler B. Big data, artificial intelligence, and structured reporting. *European Radiology Experimental*. 2018;2(1). doi:10.1186/s41747-018-0071-4

[30] Neves J, Vicente H, Esteves M, Ferraz F, Abelha A, Machado J, et al. A Deep-Big Data Approach to Health Care in the AI Age. *Mobile Networks and Applications*. 2018;23(4):1123-1128. doi:10.1007/s11036-018-1071-6

[31] Poldrack RA, Gorgolewski KJ. Making big data open: data sharing in neuroimaging. *Nat Neurosci*. 2014;17(11):1510-1517.

[32] Ding X, de Castro Caparelli E, Ross TJ. Big Data Era in Magnetic Resonance Imaging of the Human Brain. *Signal Processing and Machine Learning for Biomedical Big Data*. Published online 2018:21-54. doi:10.1201/9781351061223-3

[33] Tahmassebi A, Gandomi AH, McCann I, Schulte MH, Goudriaan AE, Meyer-Baese A. Deep learning in medical imaging: fmri big data analysis via convolutional neural networks. In: *Proceedings of the Practice and Experience on Advanced Research Computing.* ; 2018:1-4.

[34] Zhu J, Liu AA, Chen M, Tasdizen T, Su H. Special Issue on Biomedical Big Data: Understanding, Learning and Applications. *IEEE Transactions on Big Data*. 2017;3(4):375-377. doi:10.1109/tbdata.2017.2772930

[35] Wegmayr V, Aitharaju S, Buhmann J. Classification of brain MRI with big data and deep 3D convolutional neural networks. *Medical Imaging 2018: Computer-Aided Diagnosis*. Published online 2018. doi:10.1117/12.2293719

[36] Wald LL. Ultimate MRI. *Journal of Magnetic Resonance*. 2019;306:139-144. doi:10.1016/j.jmr.2019.07.016

[37] Greenspan H, van Ginneken B, Summers RM. Guest Editorial Deep Learning in Medical Imaging: Overview and Future Promise of an Exciting New Technique. *IEEE Transactions on Medical Imaging*. 2016;35(5):1153-1159. doi:10.1109/tmi.2016.2553401

[38] Ravi D, Wong C, Deligianni F, Berthelot M, Andreu-Perez J, Lo B, et al. Deep Learning for Health Informatics. *IEEE J Biomed Health Inform*. 2017;21(1):4-21.

[39] Lundervold AS, Lundervold A. An overview of deep learning in medical imaging focusing on MRI. *Z Med Phys*. 2019;29(2):102-127.

[40] Chaudhari AS, Sandino CM, Cole EK, Larson DB, Gold GE, Vasanawala SS, et al. Prospective Deployment of Deep Learning in MRI: A Framework for Important Considerations, Challenges, and Recommendations for Best Practices. *J Magn Reson Imaging*. 2021;54(2):357-371.

[41] Mazurowski MA, Buda M, Saha A, Bashir MR. Deep learning in radiology: An overview of the concepts and a survey of the state of the art with focus on MRI. *Journal of Magnetic Resonance Imaging*. 2019;49(4):939-954. doi:10.1002/jmri.26534

[42] Montagnon E, Cerny M, Cadrin-Chênevert A, Hamilton V, Derennes T, Ilinca A, et al. Deep learning workflow in radiology: a primer. *Insights Imaging*. 2020;11(1):22.

[43] Mårtensson G, Ferreira D, Granberg T, Cavallin L, Oppedal K, Padovani A, et al. The reliability of a deep learning model in clinical out-of-distribution MRI data: A multicohort study. *Medical Image Analysis*. 2020;66:101714. doi:10.1016/j.media.2020.101714

[44] Euser AM, Zoccali C, Jager KJ, Dekker FW. Cohort studies: prospective versus retrospective. *Nephron Clin Pract*. 2009;113(3):c214-c217.

[45] Kaissis G, Ziller A, Passerat-Palmbach J, Ryffel T, Usynin D, Trask A, et al. End-to-end privacy preserving deep learning on multi-institutional medical imaging. *Nature Machine*

*Intelligence*. 2021;3(6):473-484. doi:10.1038/s42256-021-00337-8

[46] Sarma KV, Harmon S, Sanford T, Roth HR, Xu Z, Tetreault J, et al. Federated learning improves site performance in multicenter deep learning without data sharing. *J Am Med Inform Assoc*. 2021;28(6):1259-1264.

[47] Sheller MJ, Anthony Reina G, Edwards B, Martin J, Bakas S. Multi-institutional Deep Learning Modeling Without Sharing Patient Data: A Feasibility Study on Brain Tumor Segmentation. *Brainlesion: Glioma, Multiple Sclerosis, Stroke and Traumatic Brain Injuries*. Published online 2019:92-104. doi:10.1007/978-3-030-11723-8_9

[48] Zbontar J, Knoll F, Sriram A, Murrell T, Huang Z, Muckley MJ, et al. fastMRI: An Open Dataset and Benchmarks for Accelerated MRI. *arXiv [csCV]*. Published online November 21, 2018. http://arxiv.org/abs/1811.08839

[49] Pawar K, Chen Z, Jon Shah N, Egan GF. Suppressing motion artefacts in MRI using an Inception-ResNet network with motion simulation augmentation. *NMR in Biomedicine*. 2022;35(4). doi:10.1002/nbm.4225

[50] Sommer K, Saalbach A, Brosch T, Hall C, Cross NM, Andre JB. Correction of Motion Artifacts Using a Multiscale Fully Convolutional Neural Network. *AJNR Am J Neuroradiol*. 2020;41(3):416-423.

[51] Duffy BA, Zhao L, Sepehrband F, Min J, Wang DJ, Shi Y, et al. Retrospective motion artifact correction of structural MRI images using deep learning improves the quality of cortical surface reconstructions. *Neuroimage*. 2021;230:117756.

[52] Muckley MJ, Ades-Aron B, Papaioannou A, Lemberskiy G, Solomon E, Lui YW, et al. Training a neural network for Gibbs and noise removal in diffusion MRI. *Magn Reson Med*. 2021;85(1):413-428.

[53] Schelb P, Wang X, Radtke JP, Wiesenfarth M, Kickingereder P, Stenzinger A, et al. Simulated clinical deployment of fully automatic deep learning for clinical prostate MRI assessment. *Eur Radiol*. 2021;31(1):302-313.

[54] Goel A, Shih G, Riyahi S, Jeph S, Dev H, Hu R, et al. Deployed Deep Learning Kidney Segmentation for Polycystic Kidney Disease MRI. *Radiol Artif Intell*. 2022;4(2):e210205.

[55] Yasaka K, Abe O. Deep learning and artificial intelligence in radiology: Current applications and future directions. *PLOS Medicine*. 2018;15(11):e1002707. doi:10.1371/journal.pmed.1002707

[56] Zhang L, Wang X, Yang D, Sanford T, Harmon S, Turkbey B, et al. Generalizing Deep Learning for Medical Image Segmentation to Unseen Domains via Deep Stacked Transformation. *IEEE Trans Med Imaging*. 2020;39(7):2531-2540.

[57] Desai A. *Meddlr: A Flexible ML Framework Built to Simplify Medical Image Reconstruction and Analysis Experimentation*. Github https://github.com/ad12/meddlr. Accessed December 23, 2022

[58] Harvey HB, Gowda V. How the FDA Regulates AI. *Acad Radiol*. 2020;27(1):58-61.

[59] The U.S. Food and Drug Administration (FDA), Health Canada, and the United Kingdom's Medicines and Healthcare products Regulatory Agency (MHRA). Good Machine Learning Practice for Medical Device Development: Guiding Principles. Published online October 2021. https://www.fda.gov/medical-devices/software-medical-device-samd/good-machine-learning-practice-medical-device-development-guiding-principles

[60] Regulation (EU) 2016/679 of the European Parliament and of the Council of 27 April 2016 on the protection of natural persons with regard to the processing of personal data and on the free movement of such data, and repealing Directive 95/46/EC (General Data Protection Regulation) (Text with EEA relevance). Published online May 4, 2016:1-88. http://data.europa.eu/eli/reg/2016/679/oj

[61] Wu E, Wu K, Daneshjou R, Ouyang D, Ho DE, Zou J. How medical AI devices are evaluated: limitations and recommendations from an analysis of FDA approvals. *Nat Med*. 2021;27(4):582-584.

[62] Center for Devices and Radiological Health. Computer-Assisted Detection Devices Applied to Radiology Images and Radiology Device Data - Premarket Notification [510(k)] Submissions. Published online September 2022. https://www.fda.gov/regulatory-information/search-fda-guidance-documents/computer-assisted-detection-devices-applied-radiology-images-and-radiology-device-data-premarket

[63] Jia X, Ren L, Cai J. Clinical implementation of AI technologies will require interpretable AI models. *Med Phys*. 2020;47(1):1-4.

[64] van der Velden BHM, Kuijf HJ, Gilhuijs KGA, Viergever MA. Explainable artificial intelligence (XAI) in deep learning-based medical image analysis. *Med Image Anal*. 2022;79:102470.

[65] Bordin V, Coluzzi D, Rivolta MW, Baselli G. Explainable AI Points to White Matter Hyperintensities for Alzheimer's Disease Identification: a Preliminary Study. *Conf Proc IEEE Eng Med Biol Soc*. 2022;2022:484-487.

[66] Bodini M, Rivolta MW, Sassi R. Opening the black box: interpretability of machine learning algorithms in electrocardiography. *Philos Trans A Math Phys Eng Sci*. 2021;379(2212):20200253.

[67] Zeiler MD, Fergus R. Visualizing and Understanding Convolutional Networks. In: *Computer Vision – ECCV 2014*. Springer International Publishing; 2014:818-833.

[68] Zhou B, Khosla A, Lapedriza A, Oliva A, Torralba A. Learning deep features for discriminative localization. In: *Proceedings of the IEEE Conference on Computer Vision and Pattern Recognition*. ; 2016:2921-2929.

[69] Natekar P, Kori A, Krishnamurthi G. Demystifying Brain Tumor Segmentation Networks: Interpretability and Uncertainty Analysis. *Front Comput Neurosci*. 2020;14:6.

[70] Windisch P, Weber P, Fürweger C, Ehret F, Kufeld M, Zwahlen D, et al. Implementation of model explainability for a basic brain tumor detection using convolutional neural networks on MRI slices. *Neuroradiology*. 2020;62(11):1515-1518.

[71] Liao L, Zhang X, Zhao F, Lou J, Wang L, Xu X, et al. Multi-Branch Deformable Convolutional Neural Network with Label Distribution Learning for Fetal Brain Age Prediction. *2020 IEEE 17th International Symposium on Biomedical Imaging (ISBI)*. Published online 2020. doi:10.1109/isbi45749.2020.9098553

[72] Ravi KS, Nandakumar G, Thomas N, Lim M, Qian E, Jimeno MM, et al. Accelerated MRI using intelligent protocolling and subject-specific denoising applied to Alzheimer's disease imaging. doi:10.1101/2022.10.24.22281473

[73] Ravi KS, Geethanath S, Quarterman P, Fung M, Vaughan JT Jr. Intelligent Protocolling for Autonomous MRI. In: *Proceedings of International Society for Magnetic Resonance in Medicine*. ; 2020.

[74] Ravi KS, Nandakumar G, Thomas N, Lim M, Qian E, Jimeno MM, et al. Intelligent denoising. *Radiology*. Published online May 2022.

[75] Karniadakis GE, Kevrekidis IG, Lu L, Perdikaris P, Wang S, Yang L. Physics-informed machine learning. *Nature Reviews Physics*. 2021;3(6):422-440. doi:10.1038/s42254-021-00314-5

[76] Fathi MF, Perez-Raya I, Baghaie A, Berg P, Janiga G, Arzani A, et al. Super-resolution and denoising of 4D-Flow MRI using physics-Informed deep neural nets. *Comput Methods Programs Biomed*. 2020;197:105729.

[77] Borges P, Sudre C, Varsavsky T, Thomas D, Drobnjak I, Ourselin S, et al. Physics-Informed Brain MRI Segmentation. *Simulation and Synthesis in Medical Imaging*. Published online 2019:100-109. doi:10.1007/978-3-030-32778-1_11

[78] Qian C, Wang Z, Zhang X, Cai Q, Kang T, Jiang B, et al. PHYSICS-INFORMED DEEP DIFFUSION MRI RECONSTRUCTION： BREAK THE BOTTLENECK OF TRAINING DATA IN ARTIFICIAL INTELLIGENCE. *arXiv [csCV]*. Published online October 20, 2022. https://arxiv.org/abs/2210.11388

[79] Weiss T, Senouf O, Vedula S, Michailovich O, Zibulevsky M, Bronstein A. PILOT: Physics-Informed Learned Optimized Trajectories for Accelerated MRI. *arXiv [csCV]*. Published online September 12, 2019. https://arxiv.org/abs/1909.05773

[80] Li S, Gong Q, Li H, Chen S, Liu Y, Ruan G, et al. Automatic location scheme of anatomical landmarks in 3D head MRI based on the scale attention hourglass network. *Comput Methods Programs Biomed*. 2022;214:106564.

[81] Hoffmann M, Turk EA, Gagoski B, Morgan L, Wighton P, Tisdall MD, et al. Rapid head-pose detection for automated slice prescription of fetal-brain MRI. *International Journal of Imaging Systems and Technology*. 2021;31(3):1136-1154. doi:10.1002/ima.22563

[82] Masutani EM, Bahrami N, Hsiao A. Deep Learning Single-Frame and Multiframe Super-Resolution for Cardiac MRI. *Radiology*. 2020;295(3):552-561.

[83] Zhao C, Dewey BE, Pham DL, Calabresi PA, Reich DS, Prince JL. SMORE: A Self-Supervised Anti-Aliasing and Super-Resolution Algorithm for MRI Using Deep Learning. *IEEE Trans Med Imaging*. 2021;40(3):805-817.

[84] Luo S, Hu J, Yang Z, Guo D, Wei H, Fu Y. A Survey on Deep Learning for Super-Resolution of Diffusion Magnetic Resonance Imaging. *Journal of Medical Imaging and Health Informatics*. 2021;11(9):2440-2449.

[85] Iglesias JE, Schleicher R, Laguna S, Billot B, Schaefer P, McKaig B, et al. Accurate super-resolution low-field brain MRI. *arXiv [csCV]*. Published online February 7, 2022. https://arxiv.org/abs/2202.03564

[86] Li Y, Sixou B, Peyrin F. A Review of the Deep Learning Methods for Medical Images Super Resolution Problems. *IRBM*. 2021;42(2):120-133. doi:10.1016/j.irbm.2020.08.004

[87] Higaki T, Nakamura Y, Tatsugami F, Nakaura T, Awai K. Improvement of image quality at CT and MRI using deep learning. *Japanese Journal of Radiology*. 2019;37(1):73-80. doi:10.1007/s11604-018-0796-2

[88] Zhao C, Carass A, Dewey BE, Prince JL. Self super-resolution for magnetic resonance images using deep networks. In: *IEEE 15th International Symposium on Biomedical Imaging (ISBI 2018)*. ; 2018:365-368.

[89] Tanabe M, Higashi M, Yonezawa T, Yamaguchi T, Iida E, Furukawa M, et al. Feasibility of high-resolution magnetic resonance imaging of the liver using deep learning reconstruction based on the deep learning denoising technique. *Magn Reson Imaging*. 2021;80:121-126.

[90] Gregory S, Cheng H, Newman S, Gan Y. HydraNet: a multi-branch convolutional neural network architecture for MRI denoising. In: Landman BA, Išgum I, eds. *Medical Imaging 2021: Image Processing*. SPIE; 2021. doi:10.1117/12.2582286

[91] Fadnavis S, Batson J, Garyfallidis E. Patch2Self: Denoising Diffusion MRI with Self-Supervised Learning. In: *34th Conference on Neural Information Processing Systems (NeurIPS 2020)*. ; 2020:16293-16303.

[92] Hernandez AG, Fau P, Rapacchi S, Wojak J, Mailleux H, Benkreira M, et al. Improving Image Quality In Low-Field MRI With Deep Learning. *2021 IEEE International Conference on Image Processing (ICIP)*. Published online 2021. doi:10.1109/icip42928.2021.9506659

[93] Manjón JV, Coupe P. MRI Denoising Using Deep Learning. In: *Patch-Based Techniques in Medical Imaging*. Springer International Publishing; 2018:12-19.

[94] Singh R, Kaur L. Noise-residue learning convolutional network model for magnetic resonance image enhancement. *Journal of Physics: Conference Series*. 2021;2089(1):012029. doi:10.1088/1742-6596/2089/1/012029

[95] Liu S, Thung KH, Qu L, Lin W, Shen D, Yap PT. Learning MRI artefact removal with unpaired data. *Nature Machine Intelligence*. 2021;3(1):60-67. doi:10.1038/s42256-020-00270-2

[96] Zhang Q, Ruan G, Yang W, Liu Y, Zhao K, Feng Q, et al. MRI Gibbs-ringing artifact reduction by means of machine learning using convolutional neural networks. *Magnetic Resonance in Medicine*. 2019;82(6):2133-2145. doi:10.1002/mrm.27894

[97] Kromrey ML, Tamada D, Johno H, Funayama S, Nagata N, Ichikawa S, et al. Reduction of

respiratory motion artifacts in gadoxetate-enhanced MR with a deep learning-based filter using convolutional neural network. *Eur Radiol*. 2020;30(11):5923-5932.

[98]   Wang G, Gong E, Banerjee S, Martin D, Tong E, Choi J, et al. Synthesize High-Quality Multi-Contrast Magnetic Resonance Imaging From Multi-Echo Acquisition Using Multi-Task Deep Generative Model. *IEEE Trans Med Imaging*. 2020;39(10):3089-3099.

[99]   Ji S, Yang D, Lee J, Choi SH, Kim H, Kang KM. Synthetic MRI: Technologies and Applications in Neuroradiology. *J Magn Reson Imaging*. 2022;55(4):1013-1025.

[100]   Gong E, Pauly JM, Wintermark M, Zaharchuk G. Deep learning enables reduced gadolinium dose for contrast-enhanced brain MRI. *J Magn Reson Imaging*. 2018;48(2):330-340.

[101]   Kleesiek J, Morshuis JN, Isensee F, Deike-Hofmann K, Paech D, Kickingereder P, et al. Can Virtual Contrast Enhancement in Brain MRI Replace Gadolinium?: A Feasibility Study. *Invest Radiol*. 2019;54(10):653-660.

[102]   Zhu B, Liu JZ, Cauley SF, Rosen BR, Rosen MS. Image reconstruction by domain-transform manifold learning. *Nature*. 2018;555(7697):487-492.

[103]   Lee D, Yoo J, Ye JC. Compressed Sensing and Parallel MRI using Deep Residual Learning. *https://scholarworks.unist.ac.kr › handlehttps://scholarworks.unist.ac.kr › handle*. Published online April 25, 2017. https://scholarworks.unist.ac.kr/handle/201301/53624. Accessed December 25, 2022

[104]   Mardani M, Gong E, Cheng JY, Vasanawala SS, Zaharchuk G, Xing L, et al. Deep Generative Adversarial Neural Networks for Compressive Sensing MRI. *IEEE Trans Med Imaging*. 2019;38(1):167-179.

[105]   Liu F, Samsonov A, Chen L, Kijowski R, Feng L. SANTIS: Sampling-Augmented Neural neTwork with Incoherent Structure for MR image reconstruction. *Magnetic Resonance in Medicine*. 2019;82(5):1890-1904. doi:10.1002/mrm.27827

[106]   Ravi KS, Geethanath S, Srinivasan G, Sharma R, Jambawalikar SR, Lignelli-Dipple A, et al. Deep learning Assisted Radiological reporT (DART). In: *Proceedings of International Society for Magnetic Resonance in Medicine 28 (2020)*. ; 2020.

[107]   Pons E, Braun LMM, Hunink MGM, Kors JA. Natural Language Processing in Radiology: A Systematic Review. *Radiology*. 2016;279(2):329-343.

[108]   Hosny A, Parmar C, Quackenbush J, Schwartz LH, Aerts HJWL. Artificial intelligence in radiology. *Nat Rev Cancer*. 2018;18(8):500-510.

[109]   Martín Noguerol T, Paulano-Godino F, Martín-Valdivia MT, Menias CO, Luna A. Strengths, Weaknesses, Opportunities, and Threats Analysis of Artificial Intelligence and Machine Learning Applications in Radiology. *J Am Coll Radiol*. 2019;16(9 Pt B):1239-1247.

[110]   Ren J, Liu PJ, Fertig E, Snoek J, Poplin R, DePristo MA, et al. Likelihood ratios for out-of-distribution detection. In: *Proceedings of the 33rd International Conference on Neural Information Processing Systems*. Curran Associates Inc.; 2019:14707-14718.

[111]   Yan W, Huang L, Xia L, Gu S, Yan F, Wang Y, et al. MRI Manufacturer Shift and Adaptation: Increasing the Generalizability of Deep Learning Segmentation for MR Images Acquired with Different Scanners. *Radiol Artif Intell*. 2020;2(4):e190195.

[112]   Mei X, Liu Z, Robson PM, Marinelli B, Huang M, Doshi A, et al. RadImageNet: An Open Radiologic Deep Learning Research Dataset for Effective Transfer Learning. *Radiol Artif Intell*. 2022;4(5):e210315.

**Figures and Tables:**

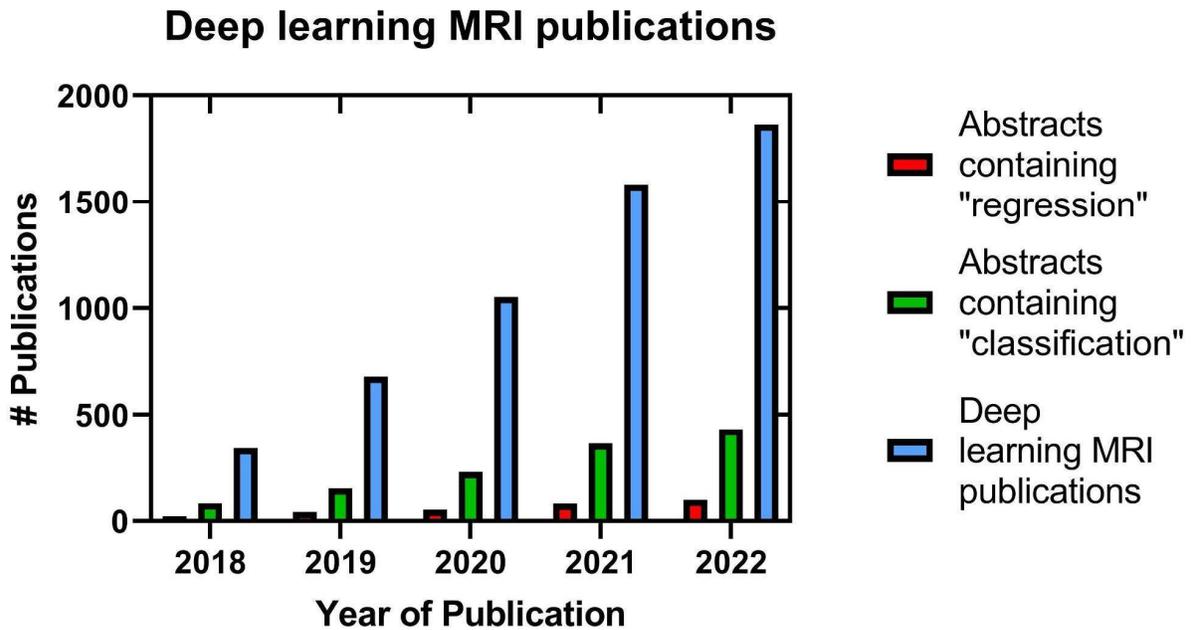

*Figure 1:* Publication trend in the past five years for deep learning in MRI. We used the PubMed database with the following keywords - regression MRI deep learning, classification MRI deep learning, and MRI deep learning.

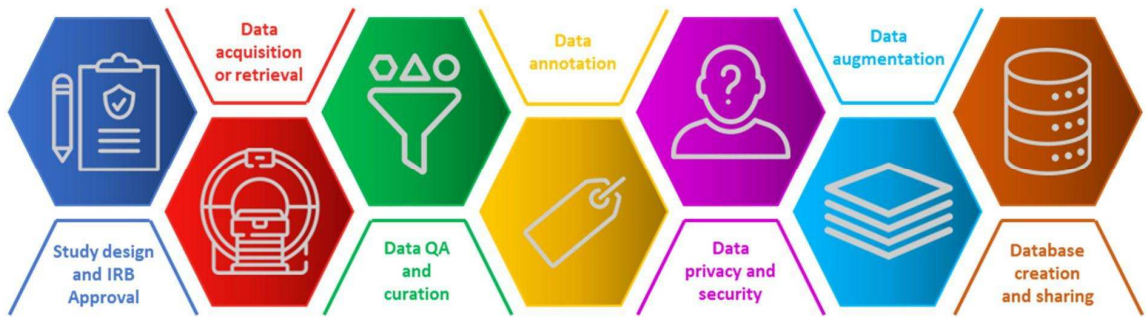

*Figure 2:* Steps for creating a dataset for the development of a deep learning model.

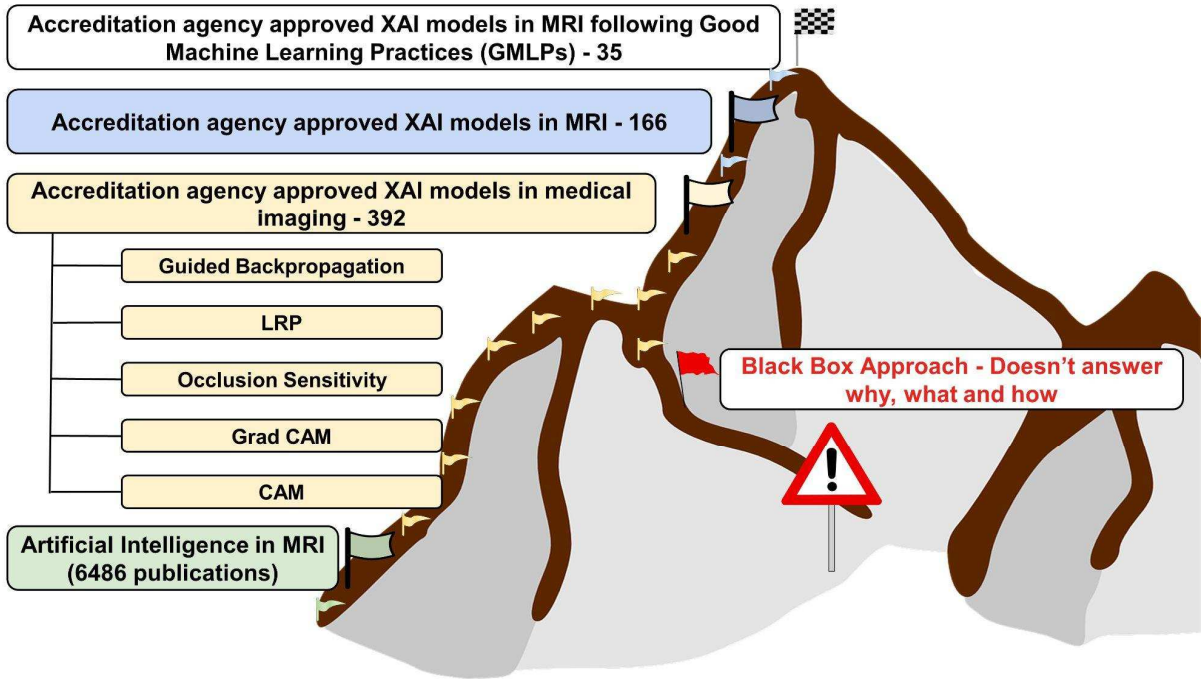

***Figure 3:*** *Mountain represents the number of publications getting reduced as we focus our scope. At the base of the mountain, we have a number of publications on AI in MRI. As we move up the mountain, we specialize more into accreditation agency approved XAI models in MRI following GMLPs. We can see that a small fraction of all publications actually make it to the top of the mountain, i.e., follow all the requirements of the accreditation agency. Many of the publications follow black box approach which doesn't explain the decision making methodology of their models and thus end up nowhere.*

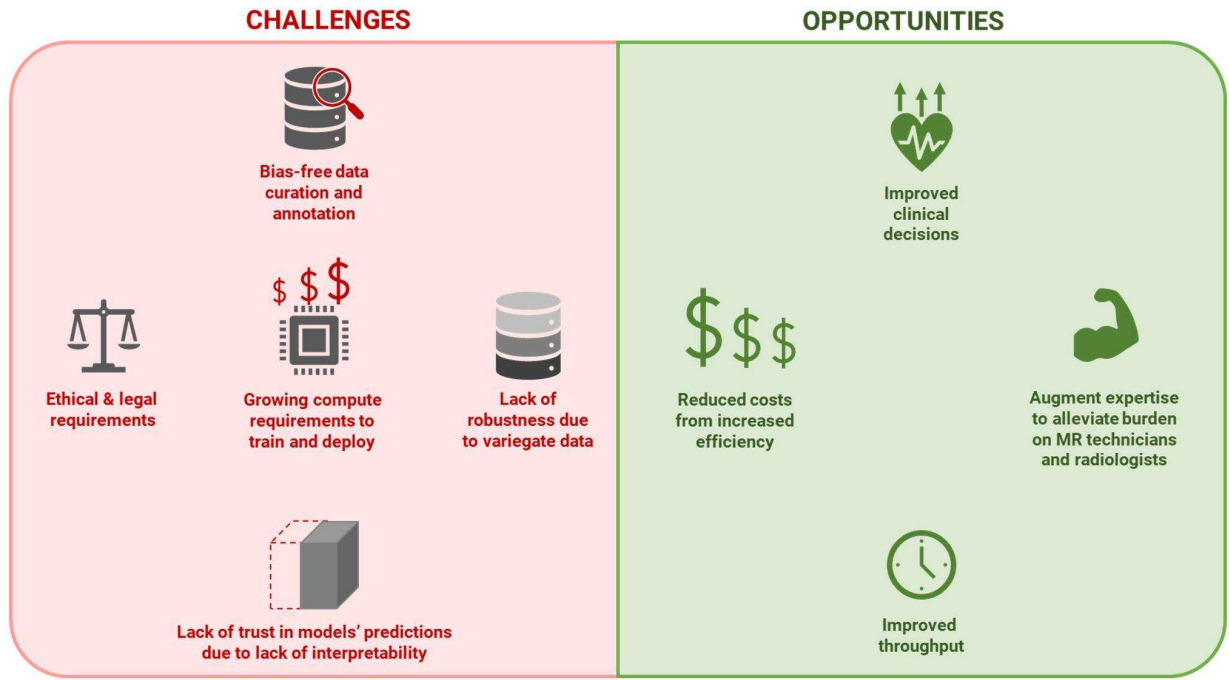

***Figure 4: Challenges and opportunities of employing deep learning (DL) in neuroimaging.***
*Developing DL methods poses several challenges, such as (clockwise from top) ensuring bias-free data curation and annotation, difficulty in assembling datasets reflective of real-world heterogeneity, black-box models stymieing development of trust in predictions, region-specific ethical and legal requirements, and managing the massive compute requirements to train and deploy models. However, unrealized opportunities such as (clockwise from top) more timely and accurate clinical decisions, augmenting available human expertise to alleviate the burden on skilled personnel, increased throughput and the associated reduction in operating costs are strong motivators to work toward a successful deployment.*

| | **Checklist of GMLPs for brain MRI** | |
|---|---|---|
| 1. | Are neuroradiologists, neuroimaging scientists, MR technician and data scientist working together throughout the whole life cycle of the product? | |
| 2. | Is the patient's personal information anonymous in the brain MR images? | |
| 3. | Is the metadata being filled for each patient scan with proper details of all parameters? | |
| 4. | Does training and testing MR datasets contain different scans? There shouldn't be any common scan in both datasets. | |
| 5. | Does reference MR dataset for validation of model have completely unique scans with same parameters as training and testing dataset? | |
| 6. | Are you using the model for segmenting brain structures from the specific contrast for which it has been trained for? If so, don't use it for other contrasts. | |
| 7. | Is the output of the model accepted and readable by the neuroradiologist? | |
| 8. | Has the model been tested in the neuroradiology department under the supervision of an expert neuroradiologist before deployment? | |
| 9. | Are the precautions and potential dangers of using the model explicitly mentioned? | |
| 10. | Is the model being updated frequently for incorporating the changes in the new scans that may occur naturally? | |

***Table 1:*** *This checklist represents the 10 guiding principles framed by the FDA for medical device development known as Good Machine Learning Practices (GMLPs). This is a simpler version of those principles rephrased specifically for experts working in brain MRI, such as neuroradiologists, neuroimaging scientists and associated data scientists. In order to get approved by the FDA, the AI model developed for MRI must fulfill all these questions.*